\documentclass[prx,aps,showpacs,twocolumn,preprintnumbers,
amsmath,amssymb,superscriptaddress,nofootinbib,longbibliography,nofootinbibjustification=justified,singlelinecheck=false]{revtex4-2}

\usepackage{xspace}
\usepackage{amsmath}
\usepackage{graphicx}
\graphicspath{{Figures/}}

\usepackage{hyperref}
\usepackage{cleveref}
\usepackage[separate-uncertainty=true]{siunitx}

\newcommand*{\algo}{\texttt{Urania}\xspace}
\newcommand*{\algobf}{\texttt{\textbf{Urania}}\xspace}
\newcommand*{\zoonum}{50\xspace}
\usepackage[svgnames]{xcolor}

\usepackage{hyperref} 
\hypersetup{
    hidelinks,
    colorlinks=true,
    breaklinks=true,
    citecolor=SteelBlue,
    filecolor=LimeGreen,
    linkcolor=MediumBlue,
    urlcolor=MediumPurple,
    pdftitle={Artificial Scientific Discovery of interferometric Gravitational Wave Detectors},
    pdfauthor={Mario Krenn, Yehonathan Drori, Rana X Adhikari}
}

\newcommand{\mpl}{Max Planck Institute for the Science of Light, Erlangen, Germany.}
\newcommand{\ligo}{LIGO Laboratory, California Institute of Technology, Pasadena, California 91125, USA.}
\begin{document}


\title{Digital Discovery of interferometric Gravitational Wave Detectors}

\author{Mario Krenn$^\S$}
\email{mario.krenn@mpl.mpg.de}
\affiliation{\mpl}
\altaffiliation{These authors contributed equally to this work.}

\author{Yehonathan Drori$^\S$}%
\email{ydrori@caltech.edu}
\affiliation{\ligo}
\altaffiliation{These authors contributed equally to this work.}

\author{Rana X Adhikari}
\email{rana@caltech.edu} 
\affiliation{\ligo}

\date{\today}

\begin{abstract}
Gravitational waves, detected a century after they were first theorized, are spacetime distortions caused by some of the most cataclysmic events in the universe, including black hole mergers and supernovae. The successful detection of these waves has been made possible by ingenious detectors designed by human experts. Beyond these successful designs, the vast space of experimental configurations remains largely unexplored, offering an exciting territory potentially rich in innovative and unconventional detection strategies. Here, we demonstrate the application of artificial intelligence (AI) to systematically explore this enormous space, revealing novel topologies for gravitational wave (GW) detectors that outperform current next-generation designs under realistic experimental constraints. Our results span a broad range of astrophysical targets, such as black hole and neutron star mergers, supernovae, and primordial GW sources. Moreover, we are able to conceptualize the initially unorthodox discovered designs, emphasizing the potential of using AI algorithms not only in discovering but also in understanding these novel topologies. We've assembled more than 50 superior solutions in a publicly available Gravitational Wave Detector Zoo which could lead to many new surprising techniques. At a bigger picture, our approach is not limited to gravitational wave detectors and can be extended to AI-driven design of experiments across diverse domains of fundamental physics.
\end{abstract}

\maketitle

\begin{center}
    \textbf{Introduction}
\end{center}
Gravitational waves are ripples in space-time produced by enormous energetic astrophysical phenomena such as the collision of two black holes or supernovae. While predicted in 1916 by Einstein as a consequence of his General Relativity~\cite{einstein1916naherungsweise}, they have only been directly observed 100 years later~\cite{abbott2016observation}. This discovery has opened a new window for the observation of phenomena in the universe, independent from electromagnetic waves, neutrinos or massive particle -- and allowing for a new branch of multi-messenger astrophysics~\cite{abbott2017multi}.

\begin{figure}
\includegraphics[width=\columnwidth]{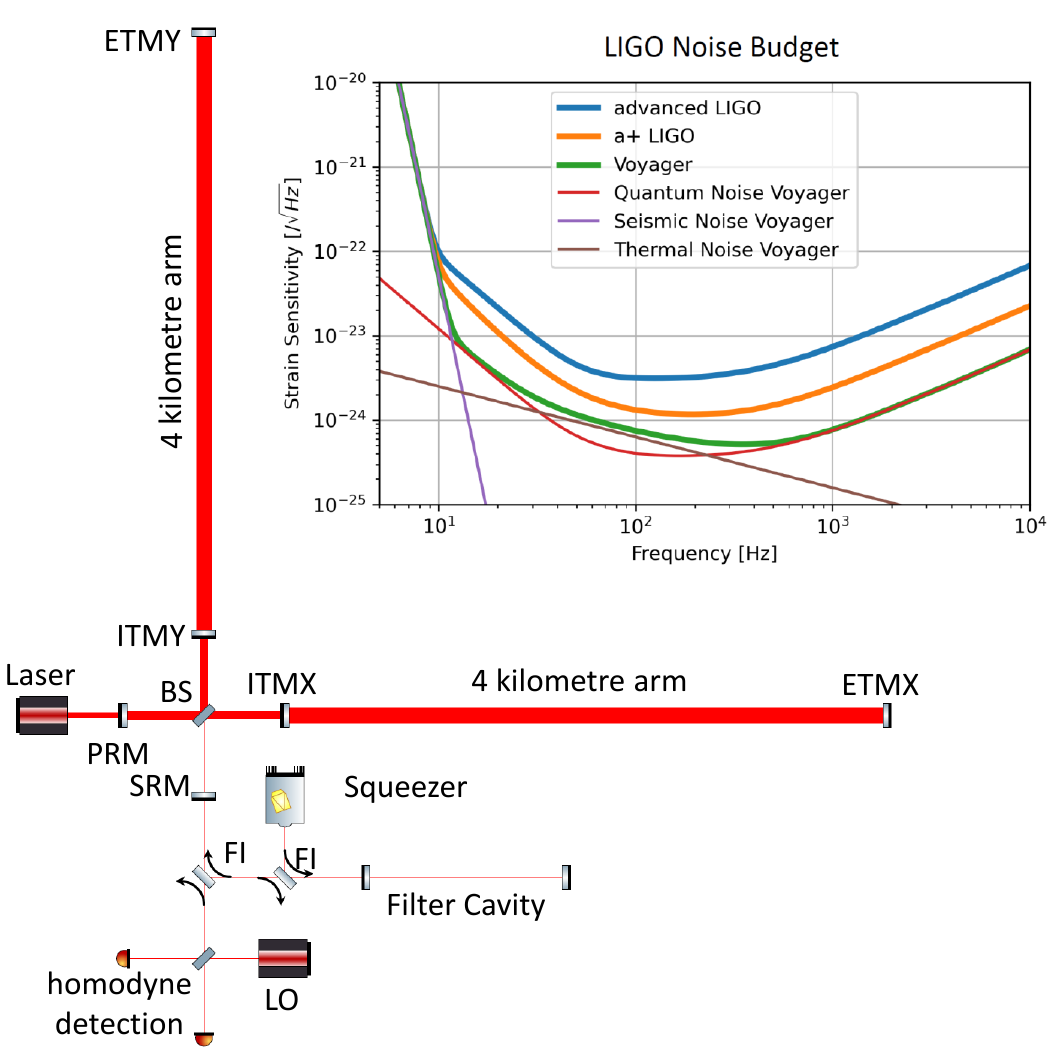}
\caption{\textbf{Setup of LIGO's GW Detector}: A laser pumps a Michelson interferometer. At the interferometer input, a power recycling mirror (PRM) recycles the light returning to the laser, increasing significantly the available photons for detection. The interferometer consists of 4\,km long arms, with a cavity in each arm (ETMX/Y and ITMX/Y). At the interferometer output, a signal-recycling mirror (SRM) reshapes the response of the interferometer. The signal is then detected with homodyne detection, by using a local oscillator (LO). Frequency-dependent squeezing is generated via a cavity and is injected into the interferometer's dark port with a Faraday isolator (FI) to reduce the dominant quantum noise in the entire detection band. The setup is similar to those of the Virgo \cite{acernese2014advanced} and KAGRA \cite{kagra2019kagra} GW detectors. Inset: The noise budget for generations of LIGO detectors: Advanced LIGO~\cite{buikema2020sensitivity}, A+ LIGO~\cite{barsotti2018a} and the next-generation Voyager detector~\cite{adhikari2020cryogenic}. The main noise contributions of the Voyager detector are quantum and classical (thermal and seismic) noise sources.}
\label{fig:aLIGODesign}
\end{figure}

The current generation of GW detectors, the Advanced Laser Interferometer Gravitational-Wave Observatory (aLIGO)~\cite{aasi2015advanced}, exploit optical interferometry, which is sensitive to the distortion of space-time. The design is based on a Michelson interferometer with multi kilometer long arms, augmented with recycling cavities and squeezed vacuum sources for reducing various noise sources. LIGO's topology and design sensitivities of current and near-future upgrades are presented in Fig.~\ref{fig:aLIGODesign}.

These detectors have all been designed by human researchers following human design principles, sometimes augmented with small-scale computational optimization of detector parameters (such as finding the ideal laser power and mirrors' reflectivities and phases)~\cite{PhysRevD.99.102004, miao2014quantum}. 

Taking a broader view -- beyond parameter optimization -- we see that there is an unimaginably large number of experimental topologies that have never been explored by human researchers. This suggests an expansive range of unexplored GW detector configurations that could not only surpass current leading designs, but could also provide entirely new and innovative experimental concepts and ideas in the field~\cite{krenn2016automated,krenn2020computer,krenn2021conceptual}.

In this manuscript, we explore the vast space of potential new topologies for GW detectors using artificial intelligence. We rephrase the discrete (and thus computationally difficult) question of topology discovery into a continuous optimization problem of a \textit{universal interferometer} (UIFO) for gravitational wave detectors. We then focus on four GW frequency regimes motivated by exciting astrophysical phenomena, such as black hole and neutron star mergers, supernovae and primordial gravitational objects. In all frequency regimes, we discover new GW topologies that outperform the current next-generation designs, under realistic experimental constraints. We further discover superior designs that mimic LIGO's current topology which are candidates for near-term upgrades. We explore and conceptualize the underlying physical principles of several of these designs by simplifying them to their core functionalities. Finally, we publish the \textit{Gravitational Wave Detector Zoo}, which contains all \zoonum diverse topologies that outperform current next-generation designs, and we hope the underlying techniques can inspire the community to develop new methods for detecting gravitational waves. Our method can not only find new superior detection schemes for rare astrophysical events, but could also be used for designing gravitational-wave based detectors for dark matter~\cite{vermeulen2021direct, hall2022advanced}, or aspects of quantum gravity~\cite{chou2017holometer, li2023interferometer}. Our work shows how advanced computational design methods can inspire new unorthodox ideas for fundamental physics experiments.

\textbf{The enormous topological search space} -- With only one laser, four optical elements (mirrors or beamsplitters) and two detectors, we can already build 3000 unique experimental topologies. With 10 elements, we could build more than 100 million unique configurations (see Appendix). Each optical element in these configurations is furthermore parameterized by continuous numbers, such as the reflectivity or the phase introduced by the element. If each element is described by only two continuous numbers, the entire search space of experiments with only 10 elements contains 100 million unique 20-dimensional search spaces. This space contains many GW detector designs invented by human researchers, including the aLIGO design. Furthermore, it contains every GW detector that has not yet been discovered. The enormous number of discrete topologies makes a systematic investigation of this space infeasible. Finding new ideas for gravitational wave detectors thus is a search problem in an extremely high-dimensional space.

\begin{figure*}
\includegraphics[width=0.98\textwidth]{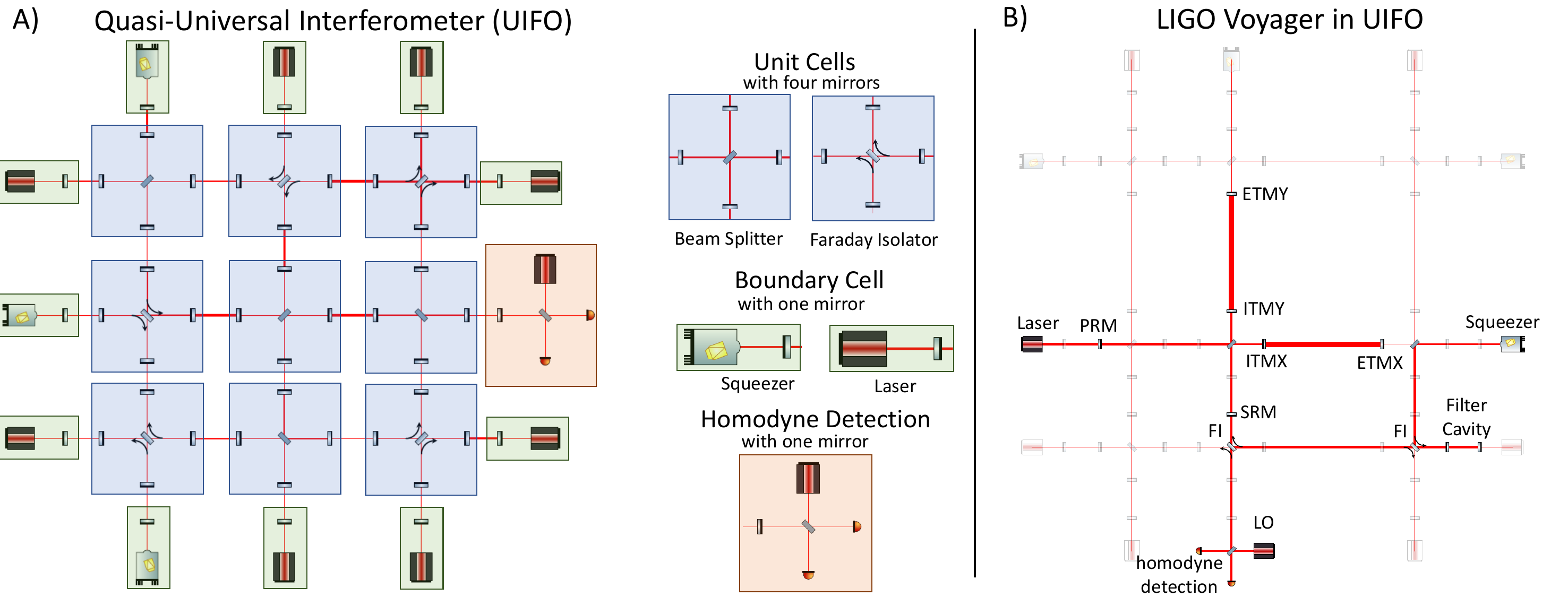}
\caption{\textbf{Details of the quasi-universal interferometer (UIFO)} A) A UIFO is a highly expressive parameterized template for an interferometer. It consists of $(n \times n)$ unit cells, light inputs at three of the four boundaries (left, up and down) and a homodyne detection cell at the right boundary. A unit cell consists of a beamsplitter or a 4-port Faraday isolator (FI) enclosed by four mirrors. The graphical representation of the FI illustrates that a light beam traveling against the direction of the arrow is transmitted, whereas a light beam traveling in the same direction as the arrow is reflected. Such 4-port routing behaviour can be realized using polarization beamsplitter with a Faraday rotator at each port. The input light can come from a laser or a squeezed vacuum source. A balanced homodyne detection scheme finally retrieves the signal. The individual elements are parametrized: Laser (power and relative phase), Squeezer (squeezing level and squeezing angle), mirror (transmission rate, relative phase, mass), beamsplitter (transmission rate, relative phase), and space (length). B) One way to inscribe LIGO's next-generation Voyager detector into UIFO. Optical elements with perfect transmission and other unused elements are rendered less visible with a semi-transparent white overlay. The thickness of the red lines indicate the power stored in the carrier frequency in a logarithmic scale. To display all the distances on the same plot, the lengths of the drawn spaces are logarithmic in the physical distances. Drawn spaces were shrunk and stretched to maintain rectangular setups.}
\label{fig:UIFO_concept}
\end{figure*}

\begin{center}
    \textbf{Digital Discovery Approach}
\end{center}
\textbf{Representing Gravitational Wave Detectors in a universal Interferometer} -- We approach this problem by reformulating the mainly discrete search problem into a continuous optimization problem, by inventing a quasi-universal interferometer (UIFO), see Fig.~\ref{fig:UIFO_concept}. The UIFO, which is inspired by the idea of universal function approximation of neural networks, is a highly expressive, parametrized optical interferometer. It consists of cells of beamsplitters (BS) enclosed by mirrors. The input into the UIFO can be either a laser or a squeezed vacuum. A balanced homodyne detection then finally retrieves the signal. The physical properties of the optical elements (such as laser power, phases, transmissivity of mirrors and beamsplitters), as well as the distances between them, are free parameters. A UIFO with $(3 \times 3)$ BS cells (UIFO$_3$) has up to 187 parameters, (UIFO$_4$ has 310, UIFO$_5$ has 463 parameters). The UIFO is constructed in such a way that setting the right parameters leads to different topological structures, e.g. we can encode the next-generation LIGO Voyager blueprint in Fig.~\ref{fig:UIFO_concept}B.

\textbf{Astrophysical design targets} -- We aim to find new designs of interferometric gravitational wave detectors for superior sensitivity in the frequency range of interesting astrophysical targets. To do so, we have to find precise parameter settings of the UIFO that lead to high-sensitivity within the desired frequency range. We search for a broadband detector in the range of 20\,--\,5000\,Hz for detections of universal sources such as binary black hole mergers, which coincides with the objective of aLIGO and the next-generation Voyager detector~\cite{aasi2015advanced, adhikari2020cryogenic}. Another target are high-sensitivity designs at low-frequency in the range of 10\,--\,30\,Hz, which is the expected signal range of  black hole mergers originating from the earliest stars in the universe~\cite{hartwig2016gravitational}. These early and distant black holes could be an interesting source of \textit{dark sirens}, for precision measurement of the Hubble constant~\cite{soares2019first}. We also search for detectors for supernov\ae\xspace explosions, which are expected in the range of 200\,--\,1000\,Hz~\cite{ott2009gravitational}. These signals have never been observed, and even the numerical and theoretical modeling of these events is still a topic of active research~\cite{andresen2019gravitational, utrobin2021supernova, vartanyan2023gravitational}. They could inform about nuclear synthesis of heavy materials~\cite{ristic2022interpolating}. Finally, we explore the range of 800\,--\,3000\,Hz, which contains expected signals from binary neutron stars mergers and postmerger physics~\cite{bernuzzi2015modeling, zhang2023gravitational}. This phenomena are to date completely unexplored, and could inform about extreme states of matter within quantum chromodynamics, including quark-gluon plasma~\cite{PhysRevD.92.023012}.

\textbf{Computationally simulating gravitational wave detectors} -- To compute the performance of a UIFO setup, we use \texttt{PyKat}~\cite{Pykat2020}, a python interface for \texttt{Finesse}. \texttt{Finesse} (\textit{Frequency domain INterfErometer Simulation SoftwarE}) is an open-source interferometer simulation program, with a main focus on gravitational wave physics~\cite{finesse}. For an experimental setup, \texttt{Finesse} computes the strain sensitivity over a frequency range of the gravitational wave, such as those in Fig.~\ref{fig:aLIGODesign}.

To computationally estimate the quality of an experimental detector design, we maximize the sensitivity while satisfying physical constraints for the parameters (such as maximal mirror reflectivity) and global constraints (such as maximal laser power that goes through an optical element). We compute the strain sensitivity at 100 discrete steps of the target frequency range (Fewer steps often lead to pathological solutions). The sensitivity is defined as the ratio between the readout noise and the interferometer optical response to strain. The optical response is calculated by modulating the spaces between the optics at a given frequency and demodulating the resulting signal at the interferometer readout. The modulation is done such that horizontal spaces are 180 degrees out-of-phase with the vertical spaces, thus mimicking the interaction of a low-frequency GW with the interferometer. The noise is calculated by propagating the quantum vacuum field, including squeezed vacuum, from all the open ports in the interferometer to the readout. Laser frequency and intensity noises are considered by calculating the amplitude and frequency transfer functions from the laser sources to the readout and projecting the intrinsic laser noises, assumed to be the same as the ones measured during aLIGO third observing run~\cite{buikema2020sensitivity}, to the readout. The parameter constraints are implemented by constraining the search space, while global design constraints are added as penalties in the computational objective function. These penalties prevent solutions with unreasonably large optical power reflected or transmitted through the optical elements and large optical power at the photon detectors. The parameter ranges of the optical elements, summarized in Table~\ref{tab:parameters}, are constrained in the same way next-generation design LIGO Voyager design, to make them experimentally feasible. 

\begin{figure}
\includegraphics[width=0.45\textwidth]{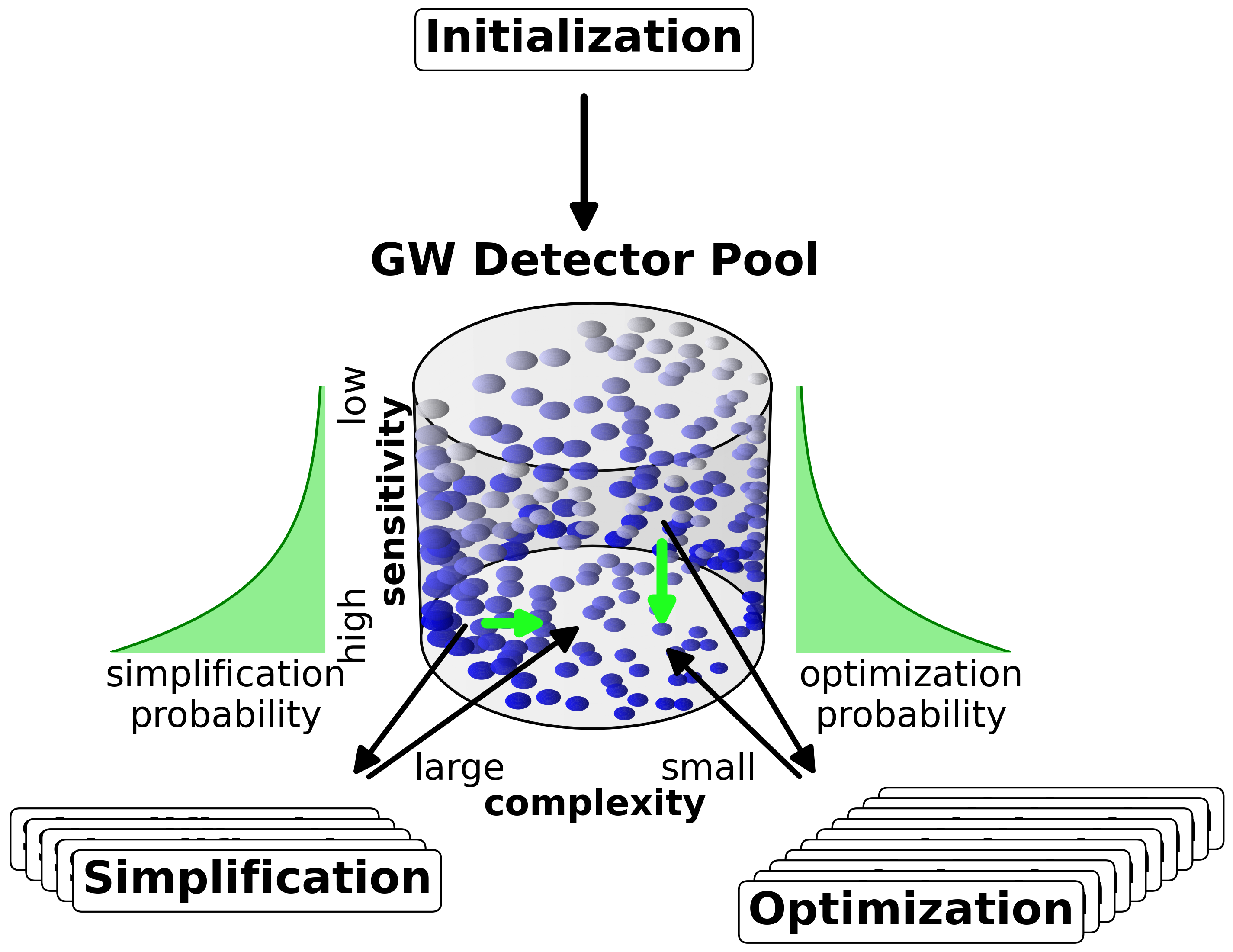}
\caption{\textbf{A sketch of the discovery engine \algo}: A pool of gravitational wave detectors is filled with random initialization of UIFOs (see Fig.~\ref{fig:UIFO_concept} for details. The horizontal location of the dots stand for complexity, and vertical for sensitivity). A large number of parallel independent optimization instances choose Boltzmann-distributed from the pool and locally optimize using numerical minimization of a loss function based on gradients and higher-order derivatives. Superior solutions replace the original instance. Subsequently, and in parallel, automated simplification algorithms choose again Boltzmann-distributed from the pool and try to reduce the complexity of the solutions. Reduced complexity solutions are added to the pool.
\label{fig:algorithm}}
\end{figure}

\begin{figure}[b]
    \centering
    \includegraphics[width=0.45\textwidth]{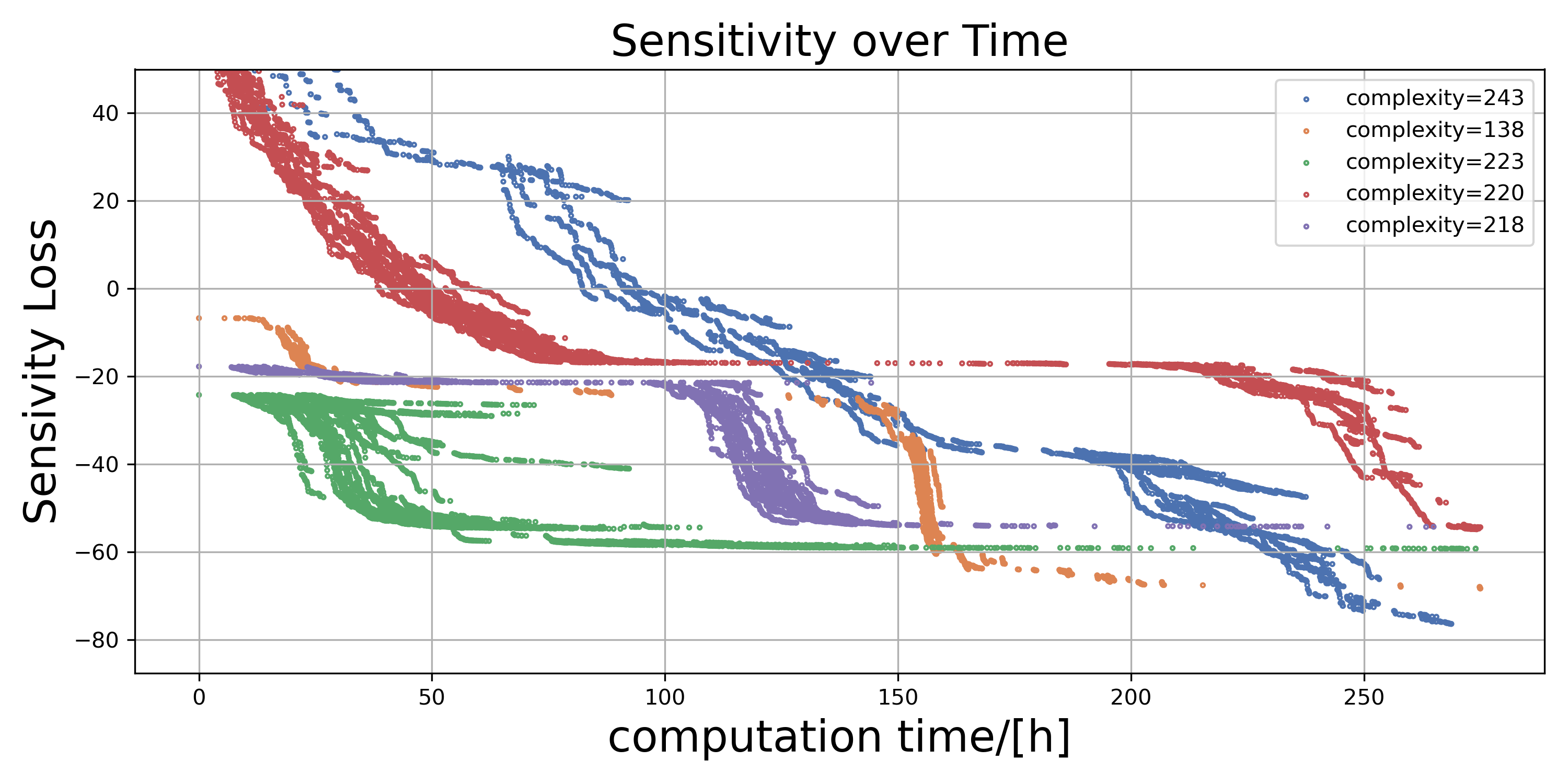}
    \caption{\textbf{Phase Transitions in Loss curves}: We show the loss evolution for (in this case, for post-merger physics targets at 2\,--\,3\,kHz) for five top-performing setups. The loss shows distinct jumps, in which the solutions discovered new abilities.}
    \label{fig:PhaseTransitions}
\end{figure}

\begin{figure*}
\includegraphics[width=0.95\textwidth]{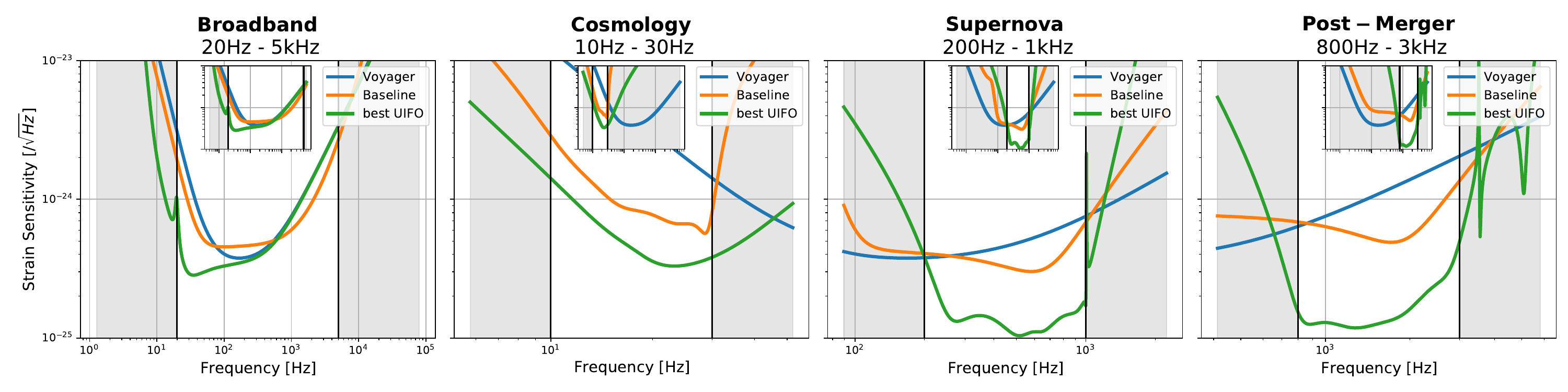}
\caption{\textbf{Strain Sensitivity curves for various objectives}: Broadband, Cosmological Targets, Supernov\ae and ranges for the post-merger analysis of binary neutron star mergers. Blue: LIGO Voyager, Orange: LIGO Voyager with parameters optimized for the specific frequency range, Green: new UIFO solutions. Noise contributions involve quantum noise, laser frequency and intensity noises. }
\label{fig:UIFO_results}
\end{figure*}

\textbf{\algobf: A parallelized hybrid local-global optimizer for scientific discovery} -- Our objective is to find experimental setups with large strain sensitivity while satisfying experimental constraints. This objective is translated into a loss function, which maximizes the detector sensitivity with experimentally feasible parameters and overall detector behaviour.

We develop \algo\footnote{In ancient Greek mythology, Urania is the muse of astronomy and stars.}, a highly parallelized hybrid local-global optimization algorithm, sketched in Fig.~\ref{fig:algorithm}. It starts from a pool of 1000s of initial conditions of the UIFO, which are either entirely random initializations or augmented with solutions from different frequency targets. \algo starts 1000 parallel local optimizations that minimize the objective function using an adapted, parameter-constrained version of the Broyden–Fletcher–Goldfarb–Shanno (BFGS) algorithm~\cite{broyden1970convergence,fletcher1970new,goldfarb1970family,shanno1970conditioning}. BFGS is a highly efficient gradient-descent optimizer that approximates the inverse Hessian matrix. For each local optimization, \algo chooses a target from the pool according to a Boltzmann distribution, which weights better performing elements in the pool higher, and adds a small noise to escape local minima. These choices add a global character to the exploration. When one of the local optimizations of \algo finds a better parameter setting for a setup in the pool, it replaces the old solution with the superior one. After a local optimization converges, \algo repeats and chooses a new target from the pool. In parallel, \algo chooses (Boltzmann distributed, thus again adding a global character) elements from the pool and simplifies it by probabilistically removing elements that have no effect on the sensitivity. These new setups have fewer parameters than their ancestors and are stored in the pool, which increases the number of setups in the pool. 

\algo is able to successfully navigate the complex search space, and identify in total \zoonum solutions that outperform the best known human-designed topologies. Surprisingly, we observe that the solutions do not continuously improve, but go through phase transitions, with periods of neglictable improvements and periods where \algo identifies a superious strategy and exploits it to its fullest -- see Fig.~\ref{fig:PhaseTransitions}. In total, we spend roughly 1.5 million CPU-hours to identify the solutions.

\begin{table}[t]
	\centering
	\renewcommand{\arraystretch}{1.5}
	\begin{tabular}{l|c|c} \hline \hline
		
		Parameter & Min & Max \\ \hline
		Optical path & \SI{10}{\centi\meter} / \SI{5}{\meter} & \SI{4}{\kilo\meter} \\
		  Suspended mass & \SI{10}{\gram} &  \SI{200}{\kilo\gram}\\
		Optical loss per element & 5\,ppm & \\
		Optical transmission& 15\,ppm & \\	
		  Squeezing level &  & \SI{10}{\decibel}\\ 
		Transmitted optical power &  & \SI{2}{\kilo\watt} \\
        Reflected optical power &  & \SI{3.5}{\mega\watt} \\   
		\hline \hline
	\end{tabular}
	\caption{Physical parameter ranges used in \algo. For UIFO optimization \SI{10}{\centi\meter} was used for lower bound on optical paths. For thermal noise minimization \SI{5}{\meter} was used. For more details see ther thermal noise section in the appendix.}
	
	\label{tab:parameters}
\end{table}

For a fair comparison with human-designed structures, we use as a baseline the next-generation LIGO Voyager detector\cite{adhikari2020cryogenic} and parametrically optimize it for each target frequency range. For that, the Voyager design in Fig.~\ref{fig:aLIGODesign} is parametrized with more than 50 variables and extensively optimized with gradient-based methods. In that way, if a solution of \algo surpasses the baseline, we know that it needs to have not only better parameters of the optical elements for the same topology but an entirely new setup topology. All comparisons take quantum noise, laser frequency and intensity noises into account. Thermal noise contributions are analysed in detail in the Appendix.

\begin{center}
    \textbf{Results}
\end{center}

We find a total of \zoonum UIFO configurations that outperform the optimized aLIGO baseline: 7 broadband solutions, 10 solutions for the cosmological window, 3 solutions for the supernova window, and the remaining 39 for analysing post-merger physics of neutron star mergers. In Fig.~\ref{fig:UIFO_results}, we show the results of the best solutions in these four frequency regimes. For these four targets we find a maximal sensitivity improvement of 4.2 compared to optimized LIGO Voyager baseline (6.8 compared the original LIGO Voyager), 2.2 (9.5), 4.0 (5.0) and 5.3 (9.0), respectively. 

The improvements of the broadband solution lie predominantly in the low-frequency regime. This would improve the observation of heavy-mass black hole mergers. It could also allow for the detection of binary neutron stars longer before they are actually merging, which can be used to send earlier signals to the detection network to identify rare multi-messenger events. 

One of the key targets of LIGO is the observation of gravitational waves from supernovae. So far, no such event has been observed. The best solution we find in this regime improves the sensitivity on average by a factor of 1.6 compared to the optimized (baseline) Voyager detector. It thereby could increase the expected observation rate by a factor of 3.8.

Finally, the post-merger solution would allow the observation of the peak of the signal of binary neutron star mergers. The expected frequency range of this peak is currently outside of the most sensitive regime of LIGO. Observing the post-merger physics could allow the analysis of the nuclear state of matter of neutron stars. The average sensitivity is improved by a factor of 4.1, potentially improving the rate by a dazzling factor of 68.7.

Details on transfer functions, thermal noise contributions, fabrication sensitivity and astrophysical observation distances are shown in the Appendix.
\begin{center}
\textbf{Physics of discovered solutions}
\end{center}

After \algo discovered the solutions, we investigate several of the top-performing ones to understand the underlying strategies that were used to go beyond human-designed detector ideas~\cite{chiang2000catching,krenn2022scientific}. We identify the core principles of the detectors such as the transducers -- the configuration which translates the gravitational wave signal to an observable optical signal, optomechanical cavities, and filter cavities. From there, we are able to identify common strategies shared by several of the solutions as well as specific individual tricks and techniques that are unique to certain detector designs (Details in the Appendix). The implementation of most of the solutions do not require extreme changes to the existing LIGO sites' infrastructure or the construction of kilometer-scale vacuum tubes, thus are candidates for potential future detector upgrades.

\begin{figure*}
    \centering
    \includegraphics[width=0.95\textwidth]{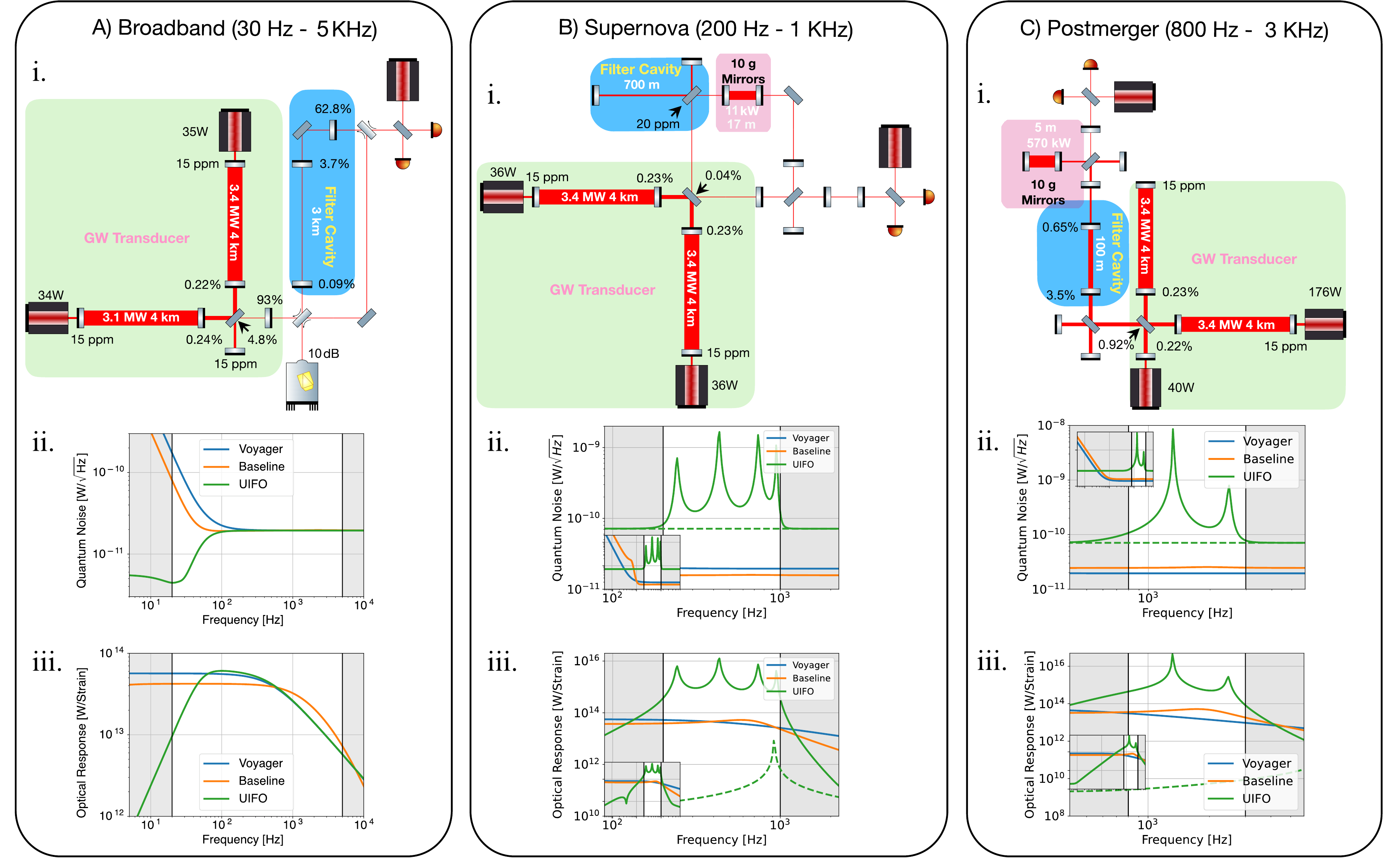}
    \caption{\textbf{Conceptualized UIFOs:} For each of the 3 simplified UIFOs \textbf{i.} Optical layout diagram illustrating the solution, highlighting essential optical parameters. The percentage alongside the mirrors and beamsplitters indicate their power transmission coefficients, while the values (in white) in or near the spaces denote their length and the optical power circulating inside them. Optical detunings are omitted for clarity. Spaces with no length notation can be arbitrarily short. As a consequence, many optical elements can be grouped together on suspended platforms, thus simplifying the control over these elements. Light traveling against the direction of the arrow of the Faraday isolator is transmitted, whereas light traveling in the same direction as the arrow is reflected. \textbf{ii.} Quantum noise of the UIFO solution (green), the baseline (orange), and Voyager (blue). \textbf{iii.} Optical response of the solution (orange), the baseline (green), and Voyager (blue). In Figures B) and C), the dashed green line shows the computed curve for the UIFO without radiation pressure in the case of the supernova and postmerger solutions. The frequency axes in B) and C) span the targets' frequency ranges. Insets show the curves in the broad frequency range.}
    
    \label{fig:ConceptualizedUIFOs}
\end{figure*}

\textbf{An unusual side-pumped L-shape GW transducer} -- The exceptional solutions in Fig.~\ref{fig:ConceptualizedUIFOs} (and in fact the vast majority of solutions we investigated), have a rather unusual GW transducer, which is different from the standard Michelson interferometer based LIGO topology. Instead of using one pump laser and a beam splitter that directs the light into the two transducer arms, these topologies use side-pumped arm cavities with a beamsplitter at the corner of an L-shaped topology. Unusual as it might look, this configuration has the same response as aLIGO, but has some interesting features. Unlike Michelson inteferometers, the arms are pumped by two lasers from the arms' high reflectivity sides, which allows the use of relatively low-power lasers. For example, as can be seen in Fig.~\ref{fig:ConceptualizedUIFOs}A, the broadband solution uses two \SI{36}{\watt} lasers while the LIGO Voyager design uses a single \SI{152}{\watt} laser. A consequence of the dual pumping is the presence of two outputs coming out of the corner beamsplitter instead of one. When the two lasers are 180$^{\circ}$ out-of-phase with each other, the carrier fields in the two outputs have opposite signs but the GW signals coming from the arms have the same sign. One can seal one of the outputs with a high-reflecting mirror to create destructive interference for the carrier and constructive interference for GW signals at the remaining output. This is implemented in the broadband solution as can be seen in Fig.~\ref{fig:ConceptualizedUIFOs}A. One can choose not to do so and use the carrier field present at the two output ports to pump optomechanical cavities for signal amplification and quantum noise squeezing, a strategy used by the other two solutions in Fig.~\ref{fig:ConceptualizedUIFOs}B and C. Finally, since the carrier and the GW signal are mixed, the corner beamsplitter serves a dual purpose as both a power-recycling and a signal-recycling mirror. This arrangement offers an advantage over a Michelson interferometer since it has one less degree of freedom that needs to be controlled. 

In practice, the dual pumping with two separate lasers requires exceptionally high-quality phase-locking, which might be challenging with current technology. However, it could still be achieved by splitting a single laser close to the corner beamsplitter and steering the split beams into the arm cavities using folding mirrors. In this case, the topology would be similar to a zero-area Sagnac~\cite{PhysRevD.38.433}, where the readout is placed behind one of the steering mirrors instead of the unused port at the input beamsplitter (more details in Appendix).

\textbf{Pondermotive squeezing} -- 
To increase sensitivity in the low frequency regime, many solutions develop a pondermotive squeezing or amplification~\cite{PhysRevD.65.042001} in the arms or in external lightweight cavities. This amplitude-dependent phase-shift originates from light hitting and changing the position of a mirror (the larger the intensity, the larger the change), which subsequently introduces a phase in the light due to the changed optical path length. In some solutions the circulating powers in the external optomechanical cavities are large and the mirror masses are small, which is a challenge for radiative cooling. As an alternative, new ideas of optical refrigeration of mirrors might be suitable~\cite{schulz2022optical}. Additionally, the development of amorphous silicon-based Optical Mirror Coatings could further dramatically improve the thermal noise budgets~\cite{PhysRevLett.120.263602,PhysRevLett.125.011102} (details in Appendix).

\textbf{The Physics of the Broadband Solution} -- 
Fig.~\ref{fig:ConceptualizedUIFOs}A shows the optical layout, quantum noise and optical response of the simplified UIFO designed for broadband frequency range. This solution was designed as an all-purpose GW detector. In this solution, GW signal exiting the open port of the \SI{4.8}{\percent} corner beamsplitter is routed through the FIs and filtered by a \SI{3}{\kilo\meter} filter cavity. A \SI{10}{\decibel} squeezed vacuum is injected into the dark port. The filter cavity allows utilization of the pondermotive squeezing generated in the arms to enhance the low-frequency sensitivity of the solution much like in the variational readout scheme proposed in~\cite{PhysRevD.65.022002}. Fig.~\ref{fig:ConceptualizedUIFOs}A \textbf{ii} shows that the quantum noise does not diverge as frequency tends to zero. Fig.~\ref{fig:ConceptualizedUIFOs}A \textbf{iii} shows that the response vanishes for zero frequency.  Additional evidence for variational readout can be found in Fig.~\ref{fig:IFOMatrix} A where it is shown that the coupling of amplitude fluctuations to the readout is highly suppressed at all frequencies. Unlike the original variational readout scheme, a small part of the light in the filter cavity is routed by a Faraday isolator back into the interferometer. This creates a weak optical spring resonance~\cite{PhysRevD.65.042001} that enhances the quantum-limited sensitivity at \SI{30}{\hertz} by a factor of 1.4 compared with the case where the filter is completely sealed.

\textbf{The Physics of the Supernova Solution} -- Fig.~\ref{fig:ConceptualizedUIFOs}B shows the optical layout and performance of the solution optimized for supernova GW signals. The GW transducer's topology is the same as the one in the broadband solution. However, in this case, the outputs of the transducer are used as inputs to a ring cavity that encompasses a \SI{17}{\meter} linear cavity formed by \SI{10}{\gram} mirrors which enhances greatly the optomechanical effects. Since the optomechanical cavity is housed inside a cavity that feeds the signal back into the interferometer, it acts as an optical spring~\cite{PhysRevD.65.042001}. In addition to the optomechanical cavity, the ring cavity also contains a \SI{700}{\meter} long filter cavity that aligns the frequency-dependant vacuum squeezing angle with the signal phase. Figs.~\ref{fig:ConceptualizedUIFOs}B \textbf{ii} and \textbf{iii} show the quantum noise and the optical response respectively. The dashed green lines in the figures show the noise and the response without radiation pressure effects (simulated by setting all the masses to infinity). Remarkably, in the presence of radiation pressure, the GW signal gets amplified by 3 orders of magnitude, while at the same time,  the quantum noise is amplified by a small amount. This is a new generalization of optical springs, that has before been applied to the standard aLIGO topology~\cite{PhysRevD.65.042001, BRAGINSKY1999241}.

\textbf{The Physics of the Postmerger Solution} --
Fig.~\ref{fig:ConceptualizedUIFOs}C shows the optical layout and performance of the solution optimized for postmerger signal of a binary neutron star coalescence. This setup is the most unintuitive of the three presented here. It is different from the other two examples presented in this paper in that, instead of pumping the IFO directly through the arm cavity, one of the lasers pumps it from the end recycling mirror. Similarly to the supernova solution, this solution uses an external short \SI{10}{\gram} cavity and a \SI{100}{\meter} filter cavity for resonantly enhancing the signal at the target frequency band. To achieve a simpler setup, there was a deliberate trade-off, leading to a reduction in sensitivity by a factor of 2 compared with the original UIFO whose sensitivity. The postmerger panel in Fig.~\ref{fig:UIFO_results} shows the sensitivity of the original UIFO.

\begin{figure}
\includegraphics[width=0.45\textwidth]{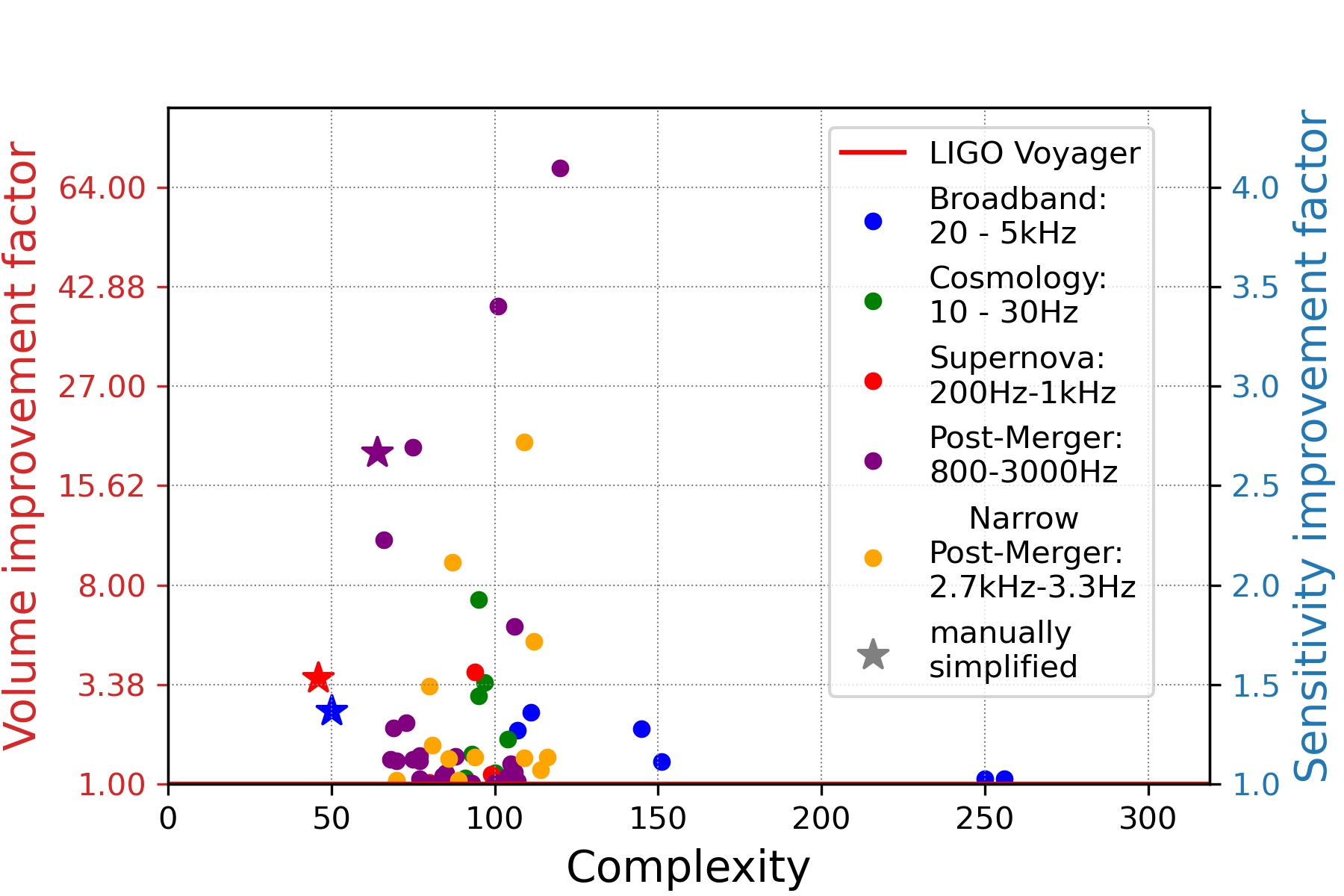}
\caption{\textbf{Gravitational Wave Detector Zoo}: We collect 59 experimental setups with improved sensitivity over the baseline, which is the parameter-optimized LIGO Voyager setup. Therefore, each solution not only has better parameters but also has a superior topology compared to LIGO Voyager. The complexity stands for the number of free parameters that remain non-trivial after simplification. The right, blue axis shows the average sensitivity improvement over the baseline. The left, red axis stands for the expected improved observational volume, based on the 3rd power of the sensitivity. The balls stand for fully automated discovered and simplified setups, the three stars are manually simplified and conceptualized setups described in Fig.~\ref{fig:ConceptualizedUIFOs}. Each of these setups has a superior topology compared to LIGO Voyager. All setups, with more information, are publicly available.
\label{fig:LIGOzoo}}
\end{figure}

\textbf{Gravitational Wave Detector Zoo} -- We collect all \zoonum solutions that outperform the baseline in a publicly available repository called \textit{Gravitational Wave Detector Zoo}~\footnote{\href{https://github.com/artificial-scientist-lab/GWDetectorZoo}{GitHub: Gravitational Wave Detector Zoo} }. Each of the solutions, by construction, has a superior topology compared to the next-generation detector LIGO Voyager. Three of the \zoonum solutions are presented in Fig.~\ref{fig:ConceptualizedUIFOs}. The examples contain five different frequency ranges, different UIFO initial conditions, variations of the UIFO detector placements, different noise contributions (such as with and without thermal and seismic noise sources), and different levels of path-length constraints. The collection of detectors might allow the community to identify new configurations, and experimental tricks and ideas that have not been discovered yet by experienced human researchers and thereby advance fundamental research of gravitational wave detectors and application of high-sensitivity (quantum) measurements in general.
\begin{center}
\textbf{Conclusions and Outlook}
\end{center}

We demonstrate how to use large-scale digital exploration to discover new superior quantum-enhanced hardware for fundamental physics research. Interperetable representations and pruning techniques allow us to conceptualize their working principles, connect them to known physical phenomena and identify new ideas or generalizations never explored by human researchers before.

This was made possible by translating a primarily discrete optimization task into a tractable continuous optimization problem. Using this approach, we were able to find novel large-scale interferometers that significantly outperform current \SI{4}{\kilo\meter} baseline GW designs at given frequency ranges without breaking near-future technological barriers. The improvements in the quantum limited sensitivity to GW signal is improved by a factor of 3-5 in the target frequency ranges compared with the Voyager design. We hope that the publication of all \zoonum superior GW detector topologies in the \textit{Gravitational Wave Detector Zoo} can inspire the community to uncover more exciting ways how to exploit the current available hardware.

Remarkably, these solutions achieve their sensitivities by utilizing single-mode, linear, time-invariant optics, free masses, Gaussian input state and homodyne detection alone. Going beyond any of these limitations might lead to more non-intuitive improvements, for example, by adding more advanced ideas from quantum technologies~\cite{miao2014quantum,schnabel2010quantum,zhuang2020distributed}, e.g. via the application of Gottesman-Kitaev-Preskill (GKP) states~\cite{grimsmo2021quantum, hastrup2022protocol} or non-Hermitian systems~\cite{lau2018fundamental}. While our solutions satisfy important experimental constraints, detailed investigations about controllability, thermal noise contributions, tolerance to noise and optical loss (including mode matching between different optical cavities will be critical. In the future, additional physical effects could be directly embedded into the optimization objectives, such as higher-order optical modes, mechanics, and thermal distortions as well as additional noise sources. Further extensions could include the co-design of cryo-systems and mechanical suspensions of mirrors.

The computational cost of more than 1\,million CPU hours stems from the expensive physical simulator. An interesting future research project could develop a auto-differentiable simulator, as commonly used for the computation of neural networks. Alternatively, an neural-network-based surrogate models could approximate the physical simulator by fast neural networks.

We want to close by remarking that our approach is not limited to GW detectors. 
A large-scale AI-based discovery framework can be successful in all cases where three criteria are met: Existence of (1) an enormously large search space, (2) a reliable and reasonably fast physical simulator, (3) a well-defined objective function. These reasonable criteria might be met for hardware discovery in many other fields of fundamental physics, for instance for dark matter~\cite{vermeulen2021direct, hall2022advanced} or dark energy detectors\cite{hamilton2015atom} or quantum gravity probes\cite{chou2017holometer, vermeulen2021experiment,li2023interferometer}. Therefore, computer-design of fundamental physics experiments might open new ways to observe the universe.
\section*{Acknowledgements}
The authors thank Daniel D. Brown and Anna C. Green for very informative discussions about the Finesse simulator, and Haixing Miao for reviewing our manuscript and useful comments that improved our presentation. The authors are thankful to the support of the Quantum Noise Working Group and the Advanced Interferometer Configurations Working Group. Y.D. and R.X.A are supported by the National Science Foundation (PHY-0823459, PHY-1764464).R.X.A. acknowledges support provided by the Institute for Quantum Information and Matter, an NSF Physics Frontiers Center (NSF Grant PHY-1733907)

\newpage

\section*{Appendix}
\subsection*{The enormous search space}
\begin{figure}[b]
\includegraphics[width=0.45\textwidth]{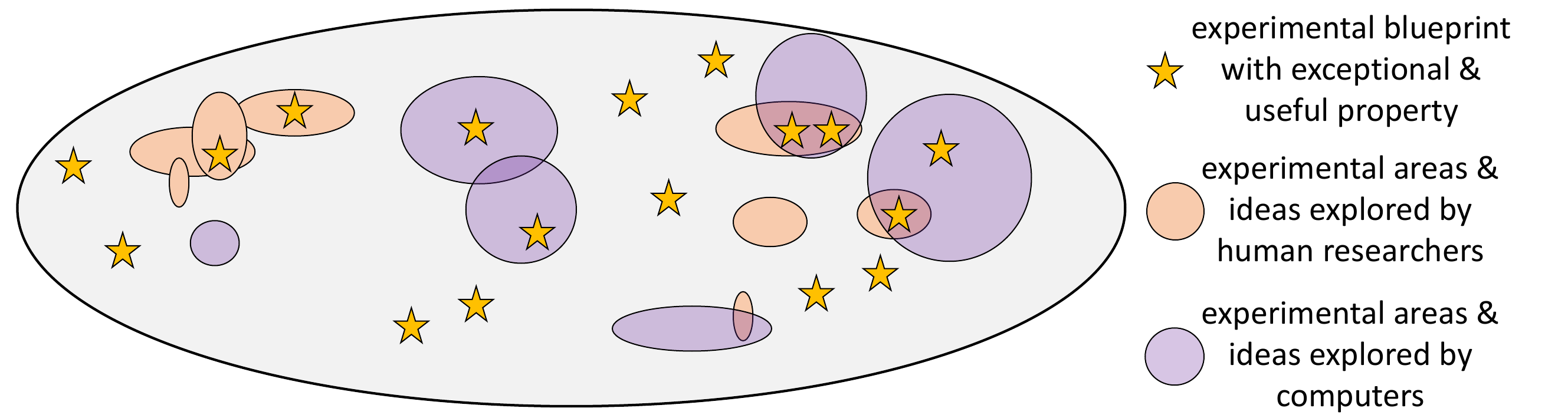}
\caption{Abstract space of all experimental configurations. This space contains all possible experimental setups, including all configurations with the exceptional property that they can detect gravitational waves (indicated by stars). In this space, creative and experienced human researchers have found numerous exciting designs (orange), such as the aLIGO design and all next-generation detector proposals. However, other useful but unorthodox designs might never be detected in this way. Here, AI-based design might help (violett).
\label{fig:SearchSpace}}
\end{figure}

\begin{figure*}
\includegraphics[width=0.95\textwidth]{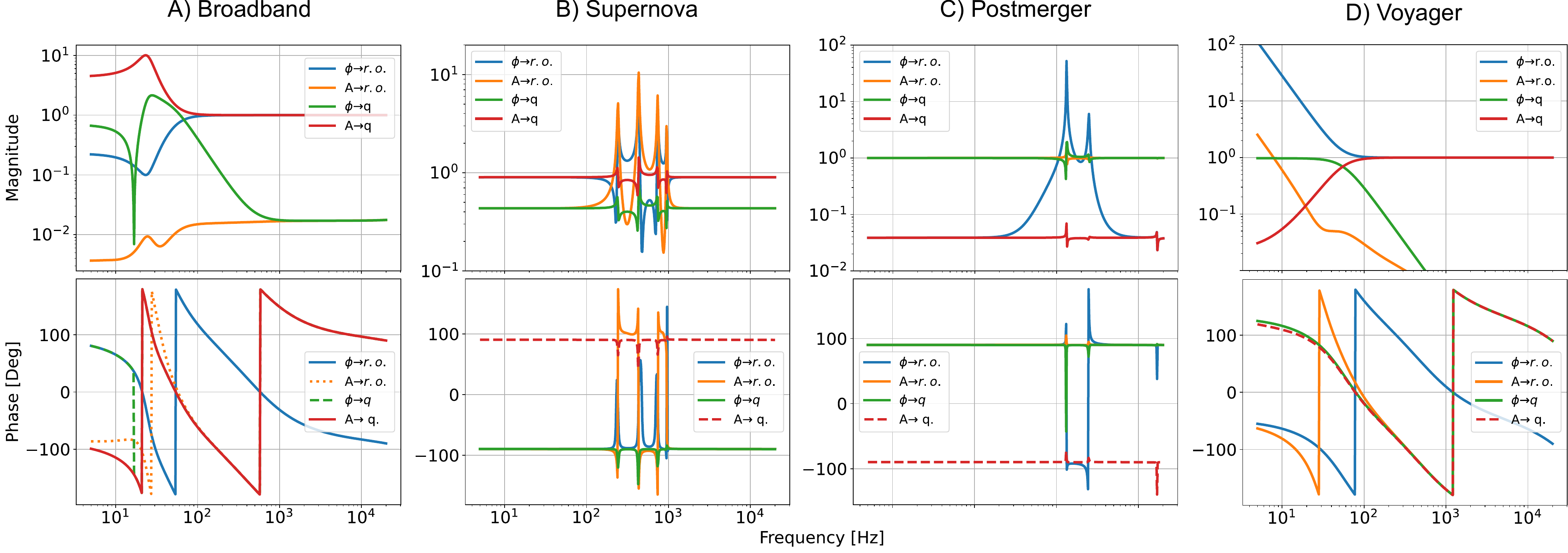}
\caption{Transfer functions between phase $\phi$ and amplitude $A$ modulations at the dark port where most the vacuum fluctuations enter, to the detector port in the nominal (r.0.) and out-of-phase (q) homodyne angles.
\label{fig:IFOMatrix}}
\end{figure*}

\noindent We estimate the size of the combinatorial search space, for experiments with one laser, and $N_{total}$ optical elements that are either beam splitters or mirrors. For simplicity, we consider our setup as a tree graph with the root node being the laser. In this simplification, we neglect setups with ring cavities or Mach-Zehnder interferometers (or combinations).

We start by computing the combinations that arise from a setup of $n$ beam splitters and $m=N_{total}-n$ mirrors. We will get three combinatorial factors that contribute. First, each beam splitter can have one out of two orientation, which leads to $b_1=2^n$ combinations. The second factor is about the location of the beam splitter. For the first beam splitter, we have only one choice (placing it inside of the laser beam). This leads to 3 paths (straight, and up and down), from which for now only two will receive light from the laser. However, any partially reflective mirror later will lead to light in this path, so we count it. Therefore, to place the second beam splitter, we have now three possible places, and create two new paths, so the next beam splitter has five choices. In total, we can place the beam splitters in $b_2=(2n-1)!!$, which is the product of all odd numbers up to $(2n-1)$.

Finally, we place the remaining $m$ mirrors. The $n$ beam splitters created $(3n+1)$ possible locations for the $m$ mirrors. This problem is directly related to the famous combinatorial \textit{Stars and Bars} problem, for which the solution is $b_3=\binom{3n+m}{3n}$. Therefore in total for a specific number $n$ of beam splitters, we get the number of $c_n=2^n (2n-1)!! \binom{3n+m}{3m}$. To get the total combination of setups with $N_{total}$ elements, we sum over all values of $n$, and get

\begin{align}
c^{N_{total}}_{Total}=\sum_{n=0}^{N_{total}} 2^n (2n-1)!! \binom{3n+m}{3m}
\end{align}

This number increases even further if we consider two detectors at the end of the setup to compute the signal. In this case, the estimation changes to

\begin{align}
d^{N_{total}}_{Total}=\sum_{n=0}^{N_{total}} 2^n (2n-1)!! \binom{3n+m}{3m} \binom{3n+1}{2}.
\end{align}

For $N_{total}=3$, we find $c^{3}_{Total}=225$ and $d^{3}_{Total}=3420$, for $N_{total}=5$, we find $c^{5}_{Total}=59,759$ and $d^{5}_{Total}=2,598,330$ and for $N_{total}=8$ we have already more than 1 billion possible configurations. In addition, each of these discrete topologies has a large continuous space that stems from the parameters of the optical components (e.g. laser power, reflectivity, phase change). Arguably, many of these setups will have the same performance, and might not be useful gravitational wave topologies.But hidden inside this enormous space, there are rare remarkable configurations, which our AI-assisted approach aims to find.

\subsection*{Details on Loss Function}

\noindent The loss function $Loss_{target}$ for discovering the targets consists of a component the strain sensitivity, and penalty for transmitting laser power through optical elements (to prevent power-induced thermal distortions), and a penalty for laser power at final detectors (to prevent bleaching of the photodetector). Formally,
\begin{equation}
Loss_{total}=-\textnormal{Loss}_{Strain} + \alpha \cdot\textnormal{Penalty}_{damage} + \beta \cdot\textnormal{Penalty}_{bleach},\nonumber
\label{eq:loss}
\end{equation}
with
\begin{align}
\textnormal{Loss}_{Strain}\approx\int_{f_0}^{f_1}\log\left(S(f)\right) df,
\end{align}
where $f_0$ and $f_1$ define the frequency interval, and $S(f)$ is the strain sensitivity (computed via Finesse~\cite{finesse}) at frequency $f$, which consists of all quantum noises, intensity, and frequency noise of the lasers. In addition, in the Gravitational Wave Detector Zoo, we demonstrate examples that outperform the LIGO Voyager baseline taking seismic and thermal noise into account, which are the main contributions at low frequencies. The effect of seismic noise is negligible for higher frequency ranges and does therefore not affect the supernovae and post-merger solutions. For the numerical computation, we discretize the interval into 100 discrete frequency steps. If we would use less, the \algo finds pathological solutions which place very narrow resonances precisely at the evaluated frequency.

Furthermore, the power damage penalty for transmitting objects ($TO$, mirrors and beamsplitters) and for photon bleaching are quasi-discrete functions (a cumulative logistic function CDF),
\begin{align}
\textnormal{Penalty}_{damage}=\sum_{TO}CDF\left(c_1 \left( p(TO) - co_p\right)\right),\\
\textnormal{Penalty}_{bleach}=\sum_{Det}CDF\left(c_2 \left( p(Det) - co_b\right)\right)
\end{align}
where $p()$ stands for the power of a transmitting object or a detector, the power cutoffs $co_p=5kW$ and $co_b=10mW$, and $\alpha$, $\beta$, $c_1$, $c_2$ are hyperparameters which are set empirically.

\begin{figure*}
\includegraphics[width=0.95\textwidth]{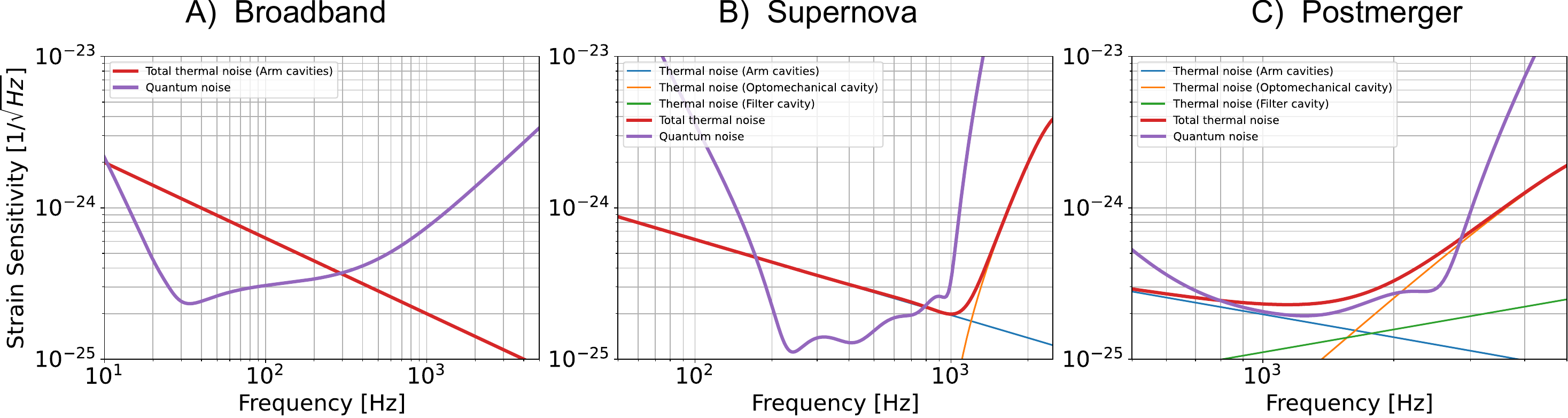}
\caption{Calculated Brownian coating thermal noise contribution of the different simplified UIFO solutions plotted together with their quantum noise. The Bragg stack of all mirrors was not optimized and was assumed to be the same as that of the end test mass of Voyager. The stability g-factor for the arm and filter cavities was assumed to be 0.9, 0.99 for the \SI{100}{\meter} filter cavity in the postmerger solution, and 0.999 for the short optomechanical cavities (well within the experimental records of $g=-0.999962$~\cite{PhysRevA.98.063833}.). All other g-values are $g=0.9$. Advanced in amorphous silicon-based Optical Mirror Coatings could further dramatically improve the thermal noise budgets~\cite{PhysRevLett.120.263602,PhysRevLett.125.011102}. Introduction of tradeoffs between stability and thermal noise (by adapting the $g$-value) will be an interesting future direction for the large scale digital discovery.
\label{fig:ThermalNoise}}
\end{figure*}

\begin{figure*}
\includegraphics[width=0.95\textwidth]{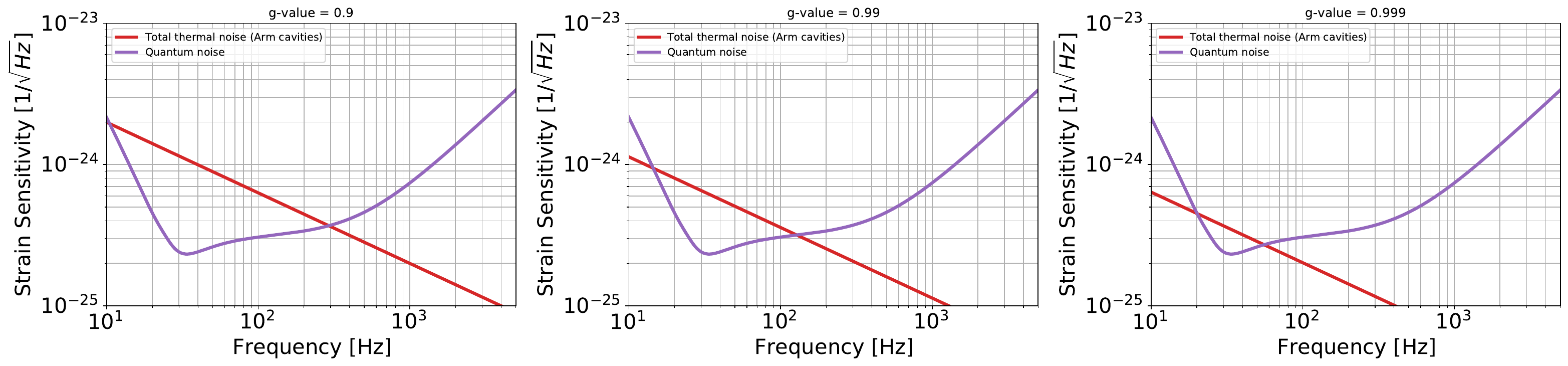}
\caption{\textbf{Brownian noise computation for the broadband solution} -- Calculated Brownian coating thermal noise for different g-values, which is a tradeoff between sensitivity and stability.
\label{fig:ThermalBroad}}
\end{figure*}

\begin{figure*}
\includegraphics[width=0.95\textwidth]{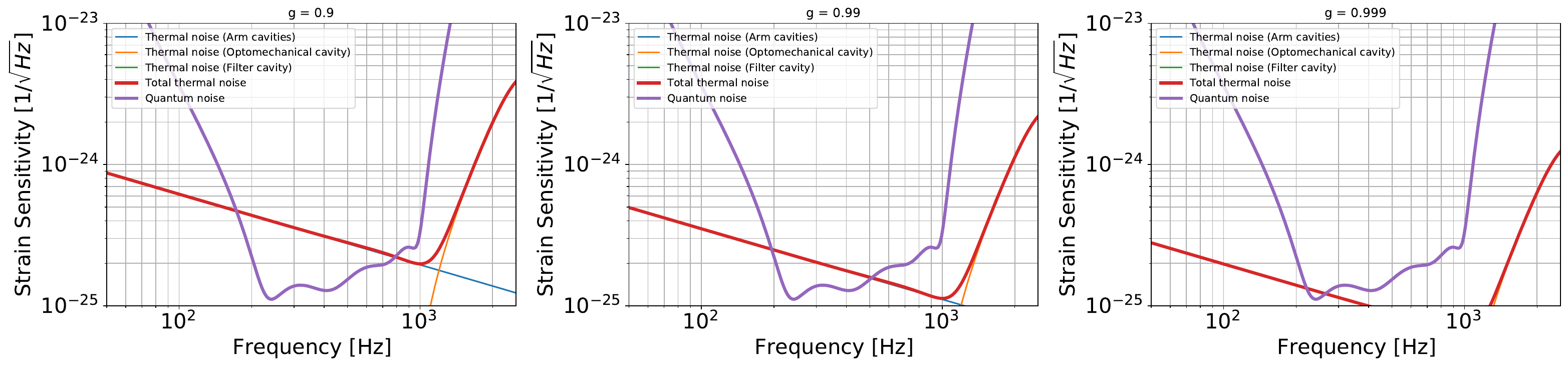}
\caption{\textbf{Brownian noise computation for Supernova solution} -- Calculated Brownian coating thermal noise for different g-values, which is a tradeoff between sensitivity and stability.
\label{fig:ThermalSupernova}}
\end{figure*}

\begin{figure}[b]
\includegraphics[width=0.48\textwidth]{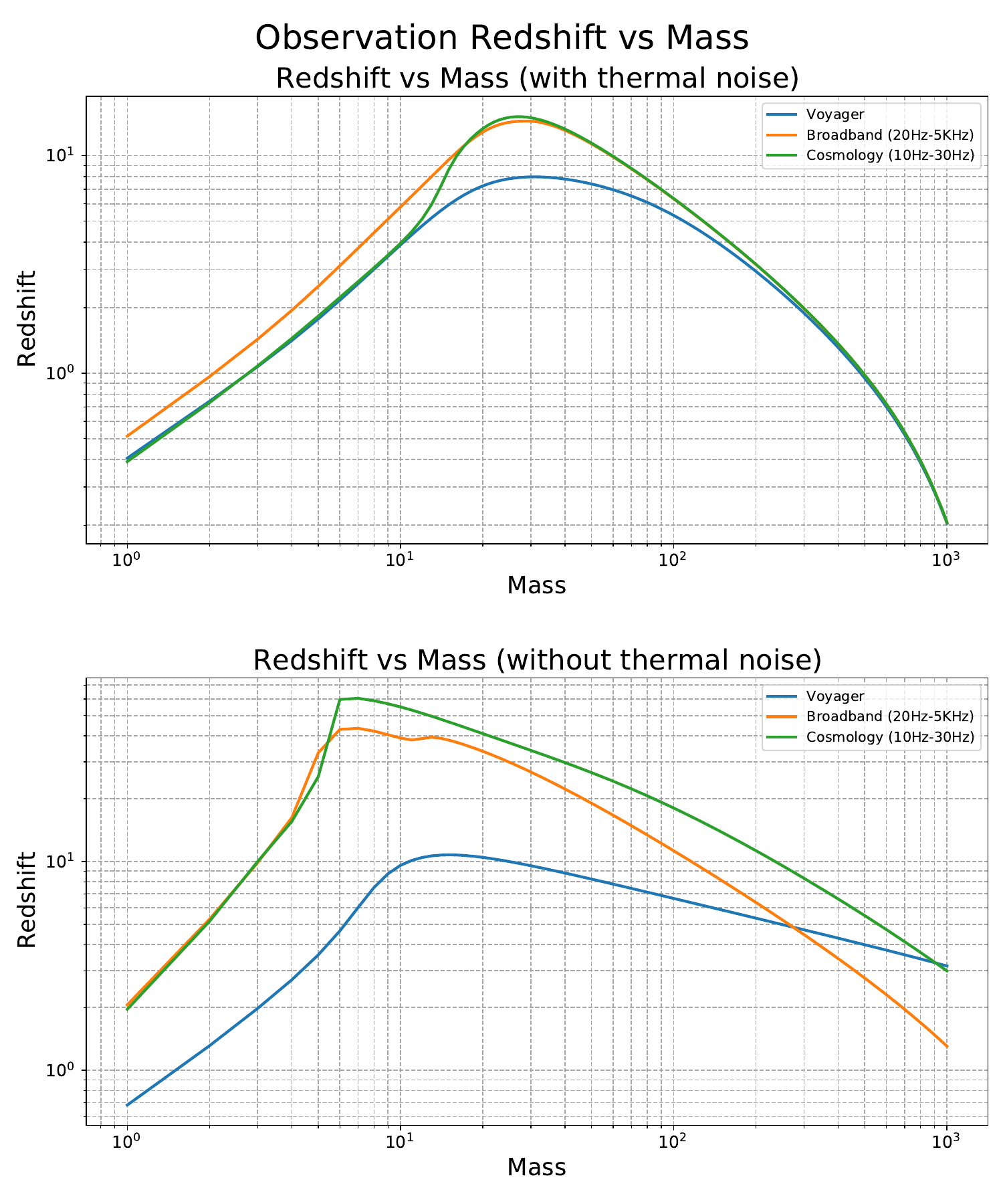}
\caption{\textbf{Redshift for observing equimassive blackhole mergers}. Masses are in units of solar masses. }
\label{fig:appendix_range_redshift}
\end{figure}

\begin{figure*}
\includegraphics[width=0.85\textwidth]{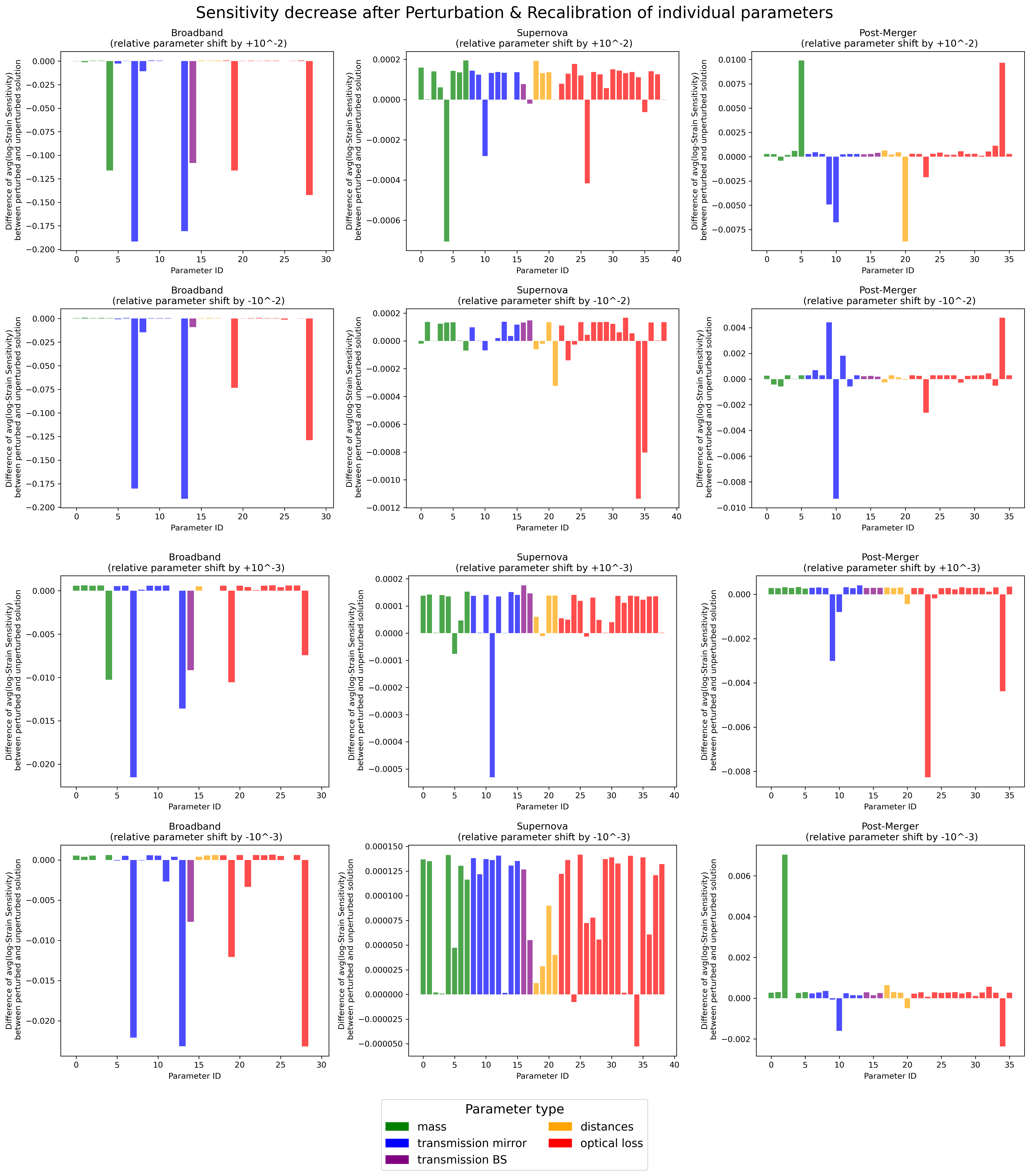}
\caption{\textbf{Change of Strain Sensitivity after Parameter Perturbation and Recalibration}. For the three UIFO in Fig.~\ref{fig:ConceptualizedUIFOs}, we compute the sensitivity of the parameters under a small perturbation. In the upper two rows, the interferometer parameters are shifted by a relative value of $10^{-2}$, in the lower two rows by $10^{-3}$. In the first and third rows, the parameters are increased by the relative value, while in the other two, they are decreased. In each case, after the perturbation, we use parameters that can be used for calibrations (such as laser and squeezing power and phases) to regain the original sensitivity. The results expressed in logs of the strain sensitivity averaged over the region of interest of the specific solution. Values larger than zero mean that the recalibration got slightly larger sensitivity than the original solution, while values smaller than zero mean that the sensitivity could not be regained perfectly after the perturbation.}
\label{fig:appendix_sensitivity}
\end{figure*}

\begin{figure*}
\includegraphics[width=0.85\textwidth]{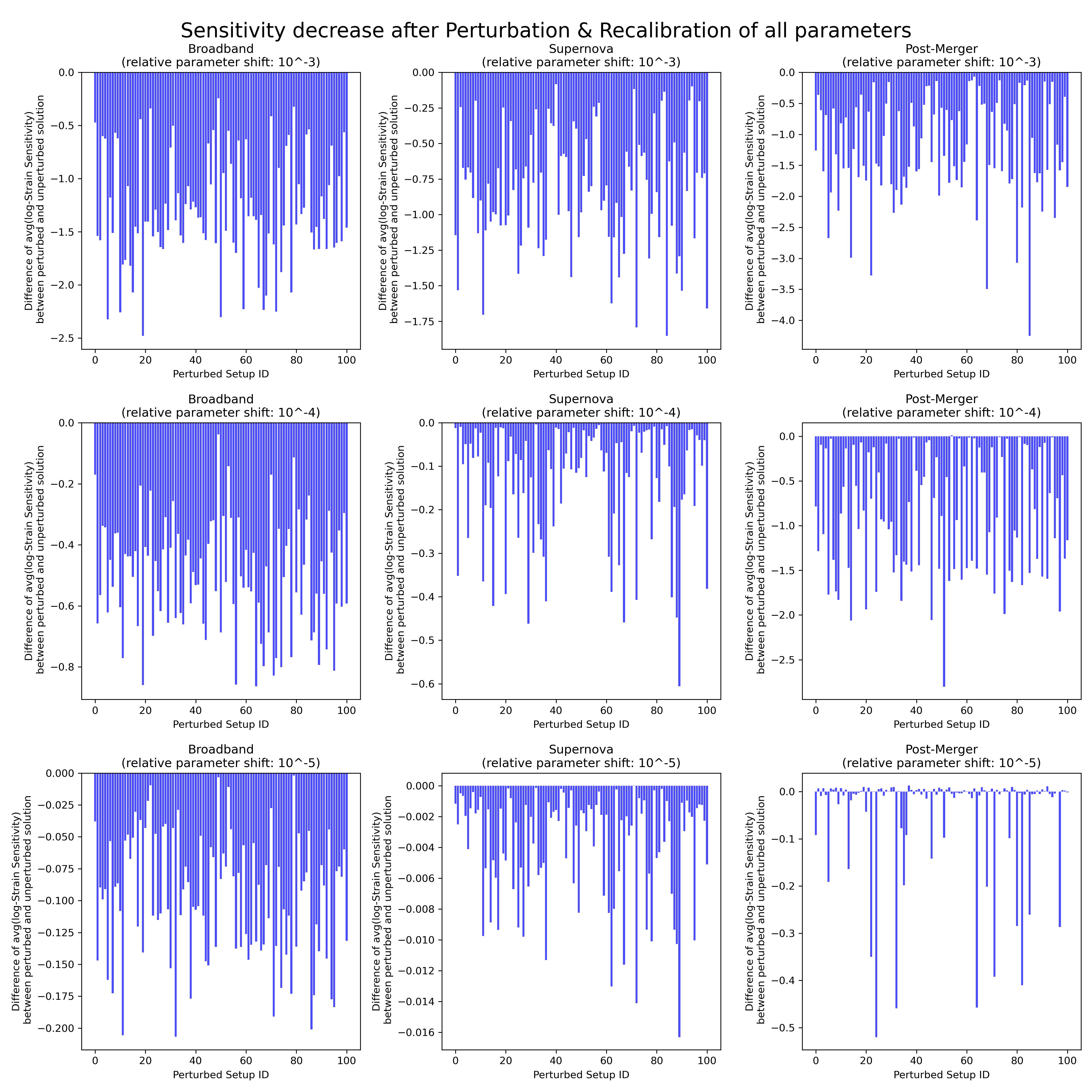}
\caption{\textbf{Change of Strain Sensitivity after random Perturbation and Recalibration of all parameters}. For the three UIFO in Fig.\ref{fig:ConceptualizedUIFOs}, we compute the sensitivity of parameters under a small perturbation. While Fig.~\ref{fig:appendix_sensitivity} shows the sensitivity for individual parameters, here we randomly perturbed all parameters. The perturbation strength is a relative value of $10^{-3}$ (first row), $10^{-4}$ (second row) and $10^{-5}$ (last row). For each of the three setups, we perform 100 random perturbations and find the best recalibration with adjustable parameters (in the same way as in Fig.~\ref{fig:appendix_sensitivity}). While the setups clearly decrease strain sensitivity, they are not unreasonably sensitive to parameter changes. }
\label{fig:appendix_many_sensitivity}
\end{figure*}

\subsection*{Simplification}
\noindent The UIFO model enables the discovery of a vast number of interferometers. However, it is far from being a practical detector due to its complexity. To bridge the gap between UIFO and real-world detectors we set out to simplify the most promising UIFO solutions. 

To simplify the UIFO, our initial step involves the removal of unnecessary optics and spaces. We accomplish this by adjusting the transmission of every optic under test to its maximum or minimum value. If this modification has a small impact on the interferometer's sensitivity, we retain the new transmission setting. Similarly, we evaluate the impact of removing spaces between optics on the interferometer's sensitivity. Spaces that have negligible effects are subsequently eliminated. This automated simplification step typically removes the majority of lasers and squeezers, leaving only a few.

Furthermore, we eliminate spaces that carry little to no signal sidebands relative to the arm cavities. We do so by replacing those spaces with laser sources having the same intensity and phase as the carrier fields in those spaces. Next, we attempt to simplify further the solutions by shortening long cavities while increasing their finesse to preserve their linewidth. Moreover, since only a handful of optics have an optimizable phase we often find solutions that contain short, low-finesse cavities that act as phase shifters. In these cases, we replace these cavities with single mirrors that possess the desired phase shift. Finally, once the majority of parameters have been eliminated, we subject the solution to another round of optimization.

Finally, after we realized our simplified solutions have unacceptably high thermal noise, we re-optimized them while raising the bound on the minimal cavity length from \SI{10}{\centi\meter} to \SI{5}{meter}. Details can be found in the thermal noise section. 

\subsection*{Input-Output relations}
\noindent The input-output relations of the solutions showcased in this paper were calculated. The transfer function between the two field quadratures at the input and output is represented by a frequency-dependant 2x2 matrix. By transforming the input quantum noise using this matrix the readout quantum noise can be calculated~\cite{PhysRevA.72.013818}. 

The computation of this matrix is done in the following way: a weak laser at carrier frequency is injected into the dark port of the interferometer. Then, using Finesse, the transfer functions from the laser amplitude and phase modulations to the readout are calculated. The transfer functions are determined at both the nominal homodyne angle and 90 out-of-phase. Finally, The matrix is normalized such that it becomes unitary at the high-frequency limit where radiation pressure effects are absent. Figure~\ref{fig:IFOMatrix} summarizes the results.

\subsection*{Thermal noise \label{sec:thermalnoise}}
\noindent The coating Brownian noise, the dominant thermal noise in cryogenic Silicon interferometers, was calculated for the simplified UIFO solutions. The thermal noise for the solutions containing short optomechanical cavities was found to be unacceptably high. To show the possibility of minimizing the thermal noise while retaining superior quantum limited sensitivity we re-optimize the simplified solutions with minimal optical path of \SI{5}{\meter}. The reason is that longer cavities allow for larger beam spot on the mirrors which in turn lower the mirror thermal noise. Figure~\ref{fig:ThermalNoise} shows the results with conservative stability criteria. The strain thermal noises from the important cavities are shown alongside the quantum noise. It was verified that the mirrors comprising those cavities contribute most of the thermal noise. The Bragg stack of the coatings was assumed to be the same as that of Voyager's end test mass~\cite{adhikari2020cryogenic}.

The thermal noise of the broadband solution and the supernova solution is dominated by the thermal noise in the arm cavities (assuming the stability factor $g$=0.9). In order to take full advantage of the low-frequency improvement of the broadband solution, improvements in the coating Brownian noise are necessary. Impressive improvements in this field were already experimentally demonstrated~\cite{PhysRevLett.120.263602,PhysRevLett.125.011102}.

For the post-merger solution, the stability factor $g$ in the arm cavities, and long filter cavities, was assumed to be 0.9, and 0.99 for the \SI{100}{\meter} filter cavity in the postmerger solution. The $g$ factor in the short optomechanical cavities was assumed to be 0.999. Recent experiment demonstrated stable operation of a short cavity with $g=-0.999962$~\cite{PhysRevA.98.063833}. Another work shows how near-unstable cavities can be experimentally demonstrated\cite{wang2018feasibility}. 

In the future, a more detailed covering of thermal contributions could lead to better solutions. In that case, a tradeoff between thermal noise and cavity stability needs to be introduced. We demonstrate different stability values for the supernova and postmerger solution in Fig.\ref{fig:ThermalBroad} and Fig.\ref{fig:ThermalSupernova}.

\subsection*{Observational redshift of solutions}
We compute the distances at which binary blackhole mergers with equal masses can be observed, see Fig.\ref{fig:appendix_range_redshift}. We plot the distance (in terms of the redshift) as a function of the black hole masses (in units of solar masses) for the Broadband solution and Cosmology solution of Fig.\ref{fig:UIFO_results}. In the upper row, we compute the distances taking into account thermal noise, which dominates at low frequencies, in the lower row, we compute the distance without thermal noise, demonstrating the potential advantage for improved thermal noise reduction. The Broadband solution outperforms LIGO Voyager mainly below 70 solar mass blackholes, while the Cosmology solution increases the distance for heavy black holes. For the computation, we use the \texttt{gwinc
inspiral-range} package, which computes several binary inspiral range measures based on the strain sensitivity of a detector~\cite{chen2021distance,inspiralrange}. 

\subsection*{Analysing non-perfect parameter settings}
The parameters in the solutions are very sensitive with respect to modifications. However, in many cases the modifications of a parameter can be re-calibrated by adjustable parameters (such as laser and squeezing parameters and phases). This raises the question of how precise the parameters of the optical components need to be produced, in particular those that are not (easily) adjusted, such as the transmissivity and optical losses of beam splitters and mirrors, the masses of optical elements, and the distances between elements.

In Fig.\ref{fig:appendix_sensitivity}, we analyze the sensitivity change for one parameter at a time, for each of the three setups in Fig.\ref{fig:ConceptualizedUIFOs} In the first two rows, we change the parameters relatively by $+10^{-2}$ and $-10^{-2}$, respectively. In the third and fourth rows, we change them by $\pm10^{-3}$. After changing the parameter, we use all adjustable parameters to re-calibrate the interferometer's sensitivity. A positive value means that the sensitivity increased compared to the unperturbed solution, while a negative value means that the sensitivity decreased. In the plot, we use the average sensitivity over the frequency region of interest for the specific solution (such as 30Hz-5KHz for the broadband solution). Specifically, we use the average of the log-strain sensitivity. That means, if a solution went from an average sensitivity of $10^{-24}$ to $10^{-23.8}$, the log-sensitivity change would be $-0.2$.

This analysis clearly shows that most of the parameter modifications can be re-calibrated with the adjustable parameters. It also demonstrated which parameters are highly sensitive and need to be produced carefully.

In a second experiment, we analyze the sensitivity of the setup to random perturbations of all parameters, followed by recalibrations using the adjustable parameters (as before, laser and squeezing powers, and phases). The results are shown in Fig.\ref{fig:appendix_many_sensitivity}, for relative perturbations on the order of $10^{-3}$, $10^{-4}$, $10^{-5}$. In a future project, the robustness of parameters towards perturbation could become a part of the loss function for the whole optimization procedure.

\subsection*{Details on the double-pumped L-shape transducer}
Unlike LIGO Voyager and the Voyager-like baseline, Fig.~\ref{fig:ConceptualizedUIFOs} shows that the quantum noise does not diverge as frequency tends to zero which is reminiscent of the characteristics of a speed meter~\cite{PhysRevD.38.433}. However, when the readout optics are removed, the solutions' responses become that of a Michelson interferometer, similar to the LIGO design. Indeed, when we replaced our transducer of the broadband solution with Voyager's we observed the same optical response as expected. This configuration is similar to the one employed by the so-called synchronous recycling interferometers, first proposed by Ronald Drever\cite{drever1983gravitational} and was recently proposed as a way to overcome the high-frequency loss limit of Michelson interferometers\cite{zhang2023gravitational}.

\bibliography{biblio}

\end{document}